\documentclass[preprint,times,sort&compress]{elsarticle}

\pdfoutput=1

\journal{ArXiv}

\usepackage{url}
\usepackage[breaklinks, hidelinks]{hyperref}

\usepackage{graphicx}

\usepackage{amsmath}
\usepackage{mathtools}

\usepackage{booktabs}

\usepackage[caption=false,font=footnotesize]{subfig}

\usepackage[acronym]{glossaries}
\setacronymstyle{long-short}
\newacronym{FPGA}{FPGA}{Field-Programmable Gate Array}
\newacronym{QoS}{QoS}{Quality of Service}
\newacronym{QoE}{QoE}{Quality of Experience}
\newacronym{OTT}{OTT}{Over-The-Top}
\newacronym{FR}{FR}{Full-Reference}
\newacronym{RR}{RR}{Reduced-Reference}
\newacronym{NR}{NR}{No-Reference}
\newacronym{ITU-T}{ITU-T}{International Telecommunication Union -Telecommunication Standardization Sector}
\newacronym{MOS}{MOS}{Mean Opinion Score}
\newacronym{KPI}{KPI}{Key Performance Indicator}
\newacronym{SG12}{SG12}{ITU-T Study Group 12}
\newacronym{IPTV}{IPTV}{Internet Protocol Television}
\newacronym{P.NAMS}{P.NAMS}{non-intrusive parametric model for assessment of performance of multimedia streaming}
\newacronym{P.NBAMS}{P.NBAMS}{non-intrusive bit-stream model for assessment of performance of multimedia streaming}
\newacronym{MOAVI}{MOAVI}{monitoring of audio-visual quality by key indicators}
\newacronym{HLL}{HLL}{High Level Language}
\newacronym{GPP}{GPP}{General Purpose Processor}
\newacronym{PSP}{PSP}{Platform Support Package}
\newacronym{GCC}{GCC}{GNU Compiler Collection}
\newacronym{FIFO}{FIFO}{First In, First Out}

\usepackage[binary-units=true]{siunitx}

\usepackage{listings}
\usepackage[T1]{fontenc}

\begin{document}

\begin{frontmatter}

\title{FPGA Implementation of the Procedures for Video Quality Assessment}

\author[aghaddress,cyfronetaddress]{Maciej Wielgosz}
\ead{wielgosz@agh.edu.pl}

\author[aghaddress,cyfronetaddress]{Micha\l{} Karwatowski}
\ead{mkarwat@agh.edu.pl}

\author[aghaddress,cyfronetaddress]{Marcin Pietro\'n}
\ead{pietron@agh.edu.pl}

\author[aghaddress,cyfronetaddress]{Kazimierz Wiatr}
\ead{wiatr@agh.edu.pl}

\address[aghaddress]{AGH University of Science and Technology, Krak\'ow, Poland}
\address[cyfronetaddress]{Academic Computer Centre CYFRONET, Krak\'ow, Poland}

\begin{abstract}
Video resolutions used in variety of media are constantly rising. While 
manufacturers struggle to perfect their screens it is also important to ensure 
high quality of displayed image. Overall quality can be measured using Mean 
Opinion Score (MOS). Video quality can be affected by miscellaneous artifacts, 
appearing at every stage of video creation and transmission. In this paper, we 
present a solution to calculate four distinct video quality metrics that can be 
applied to a real time video quality assessment system. Our assessment module 
is capable of processing 8K resolution in real time set at the level of 
30 frames per second. Throughput of 2.19 GB/s surpasses performance of pure
software solutions. To concentrate on architectural optimization, the module 
was created using high level language.
\end{abstract}


\end{frontmatter}

\section{Introduction}
\label{intro}

Nowadays, in addition to traditional \gls{QoS}, \gls{QoE} poses a real challenge for Internet audiovisual service providers, broadcasters and new \gls{OTT} services. The churn effect is linked to \gls{QoE} impact; the end-user satisfaction is a real added value in this competition. However, \gls{QoE} tools should be proactive and innovative solutions that are well adapted to new audiovisual technologies. Therefore, objective audiovisual metrics are frequently dedicated to monitoring, troubleshooting, investigating, and setting benchmarks of content applications working in real-time or off-line.

The so called \gls{FR}, \gls{RR} and \gls{NR} quality metrics are used for models standardized according to \gls{ITU-T} Recommendations. Most of the models have some limitations as 
they were usually validated using one of the following hypotheses: frame freezes 
last up to two seconds; there is no degradation at the beginning or at the end of the 
video sequence; there are no skipped frames; video reference is clean (no spatial or 
temporal distortions); there is minimum delay supported between video reference and 
video (sometimes with constant delay); and up or down-scaling operations are not 
always taken into account \cite{muscle}.

\begin{table}
\caption{The history regarding ITU-T Recommendations (based on\cite{muscle}).\label{tab:history}}
\centering
\begin{tabular}{cccc} \toprule
Model Type & Format & Rec. & Year \\ \midrule
\gls{FR} & SD	& J.144 \cite{J.144} & 2004\\
\gls{FR} & QCIF--VGA & J.247 \cite{J.247} & 2008\\
\gls{RR} & QCIF--VGA & J.246 \cite{J.246} & 2008\\
\gls{FR} & SD & J.144 \cite{J.144} & 2004\\
\gls{RR} & SD & J.249	\cite{J.249} & 2010\\
\gls{FR} & HD & J.341 \cite{J.341} & 2011\\
\gls{RR} & HD	& J.342	\cite{J.342} & 2011\\
Bitstream & VGA--HD & In progress & Exp. 2014\\
Hybrid & VGA--HD & In progress & Exp. 2014\\ \bottomrule
\end{tabular}
\end{table}

\begin{table}
\caption{Synthesis of \gls{FR}, \gls{RR} and \gls{NR} \gls{MOS} models (based on\cite{muscle,Emmanuel}).\label{tab:synthesis}}
\centering
\begin{tabular}{ccccc} \toprule
& & FR & RR & NR \\ \midrule
5*Resolution & HDTV & J.341 \cite{J.341} & n/a   & n/a  \\
& SDTV & J.144 \cite{J.144} & n/a   & n/a  \\
& VGA & J.247 \cite{J.247} & J.246 \cite{J.246} & n/a  \\
& CIF & J.247 \cite{J.247} & J.246 \cite{J.246} & n/a  \\
& QCIF & J.247 \cite{J.247} & J.246 \cite{J.246} & n/a\\ \bottomrule
\end{tabular}
\end{table}

In the past, metrics based on three historical video artifacts (blockiness, 
jerkiness, blur) were sufficient to provide an efficient predictive result. 
Consequently, most models are based on measuring these artifacts for producing a 
predictive \gls{MOS}. In other words, the majority of the 
algorithms generating the predicted \gls{MOS} show a mix of blur, blockiness, and 
jerkiness metrics. The weighting between each of these \glspl{KPI} could be a simple mathematical function. If one of the \glspl{KPI} is 
not correct, the global predictive score is completely wrong. Other \glspl{KPI} are 
usually not taken into account (exposure time distortion, interlacing, etc.) in 
predicting \gls{MOS} \cite{muscle}.

The \gls{ITU-T} has been working on \gls{KPI}-like distortions for many years (please refer 
to \cite{P.930} for more information). The history of the recommendations is 
shown in Tab.~\ref{tab:history}, while metrics based on video signal only are 
shown in Tab.~\ref{tab:synthesis}, both based on \cite{muscle}.

Related research in \cite{4711777} addresses measuring multimedia quality in 
mobile networks with an objective parametric model \cite{muscle}.

\Gls{SG12} is currently working on modeling standards for multimedia and \gls{IPTV} based on bit-stream information. Q14/12 work group is responsible for the projects provisionally known as \gls{P.NAMS} and \gls{P.NBAMS}\cite{muscle}. 

\Gls{P.NAMS} utilizes packet-header information (e.g., from IP through MPEG2-TS), 
while \gls{P.NBAMS} also uses the payload information (i.e., coded 
bit-stream)\cite{Takahashi_Yamagishi_Kawaguti_2008}. 
However, this work focuses on the overall quality (in \gls{MOS} units), while 
\gls{MOAVI} is focused on \glspl{KPI}\cite{muscle}.

Most of the recommended models are based on global quality evaluation of video 
sequences as  in the \gls{P.NAMS} and \gls{P.NBAMS} projects. The predictive score is 
correlated to subjective scores obtained with global evaluation methodologies 
(SAMVIQ, DSCQS, ACR, etc.). Generally, the duration of video sequences is 
limited to \num{10} or \SI{15}{\second} in order to avoid the forgiveness effect 
(the observer is unable to score the video properly after \SI{30}{\second}, and may 
give more weight to artifacts occurring at the end of the sequence). When one 
model is deployed for monitoring video services, the global scores are provided 
for fixed temporal windows and without any acknowledgment of the previous 
scores\cite{muscle}.

Generally, the time needed to process such metrics is long even when a powerful 
machine is used. Hence, measurement periods have been short and never extended 
to longer periods. As a result, the measurements miss sporadic and erratic 
audiovisual artifacts. 

The concept proposed here, partly based on the framework for the integrated 
video quality assessment published in \cite{ISI:000309861700014}, is able to 
isolate and focus investigation, set up algorithms, increase the monitoring 
period and guarantee better prediction. Depending on the technologies used in 
audiovisual services, the impact of \gls{QoE} can change completely. The scores are 
separated for each algorithm and preselected before the testing phase. Then, 
each \gls{KPI} can be analyzed by working on the spatially and/or temporally 
perceived 
axes. The classical metric cannot provide pertinent predictive scores with 
certain new audiovisual artifacts such as exposure distortions. Moreover, it is 
important to detect the artifacts as well as the experience described and 
detected by the consumers. In real-life situations, when video quality of audiovisual services decreases, the customers can call a helpline and describe the annoyance and visibility problems; they are not required to provide a \gls{MOS}.

There are many possible reasons for video disturbance, and they can arise at 
any point along the video chain transmission (filming stage to end-user stage).
The main concern of the authors of the papers is an efficient hardware 
implementation of proposed 
solution. This is addressed using hardware development techniques decreasing 
latency and throughput of the system which 
is a challenging task partially covered in the following papers 
\cite{wielgosz0,wielgosz1,wielgosz2,wielgosz3,wielgosz4}.

\section{Related work}
\label{RelatedWork}

Automated video quality assessment has been an issue addressed in many papers 
in previous years. Ligang Lu et al. in paper \cite{Ligang} presented 
a no-reference solution for MPEG video stream measuring quantization error and 
blocking effect. Their solution showed positive correlation with other methods. 
However, because of the technology available at the time of publication, 
their system throughput is far from modern requirements. 
Marcelo de Oliveira et al. \cite{Oliveira} successfully implemented 
Levenberg-Marquardt method in low end platforms using VHDL. They showed that 
hardware implementation results maintain a strong correlation with software 
solution, despite reduced precision due to usage of fixed point arithmetic.
Neborovski et al. \cite{Neborovski} implemented field-offset detection, 
blurring and ringing measurements in \gls{FPGA}. Their language of choice was Verilog, using 
platform based on Virtex~4 they achieved real time processing for fullHD 
resolutions.

\section{Video quality assessment}
\label{VideoQualityAssessment}

\begin{table}[b]
\caption{Notation used in equations.\label{tab:notation}}
\centering
\begin{tabular}{ll} \toprule
Symbol & Description \\ \midrule
$\mathit{BLX}$ & number of horizontal blocks in a frame\\
$\mathit{BLY}$ & number of vertical blocks in a frame\\
$\mathit{sortMeanBL}$ & ordered sequence of the average luminance of blocks\\
$\mathit{sortSumBL}$ & ordered sequence of the luminance sums calculated for each block\\
\bottomrule
\end{tabular}
\end{table}
 
This paper addresses a challenging task of building a module capable of 
accelerating the metrics computations. 
Consequently, the designed module produces video quality assessment in  
real time for each video frame.
The selected four metrics were implemented in hardware:

\begin{enumerate}[(ii)]
 \item blockiness
 \item exposure
 \item blackout
 \item interlace
\end{enumerate}

The choice of the metrics was driven by their performance and hardware 
implementation feasibility. 

The authors designed and implemented a single module for all the four metrics. 
Such an approach enables hardware units sharing among the metrics architectures 
and it boosts the overall throughput of the video assessment quality module. 

Blockiness and the exposure metrics are presented in \cite{6181021,6011903}, 
respectively.

This section presents an overview of all the metrics and the algorithms used in this work. Notation used in equations is presented in Tab.~\ref{tab:notation}.

\subsection{Blocking}

\begin{figure}
\begin{center}
\includegraphics[width=0.72\hsize]{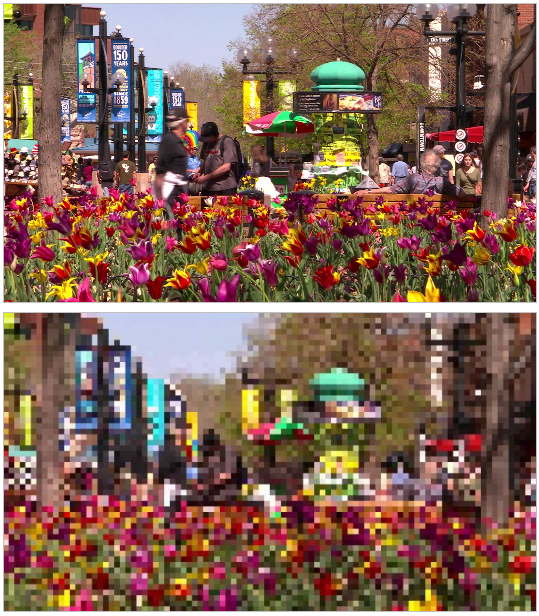} 
\caption{Blockiness artifact}
\label{fig:blockiness0}
\end{center}
\end{figure}

\begin{figure}
\centering
\includegraphics[width=0.5\hsize]{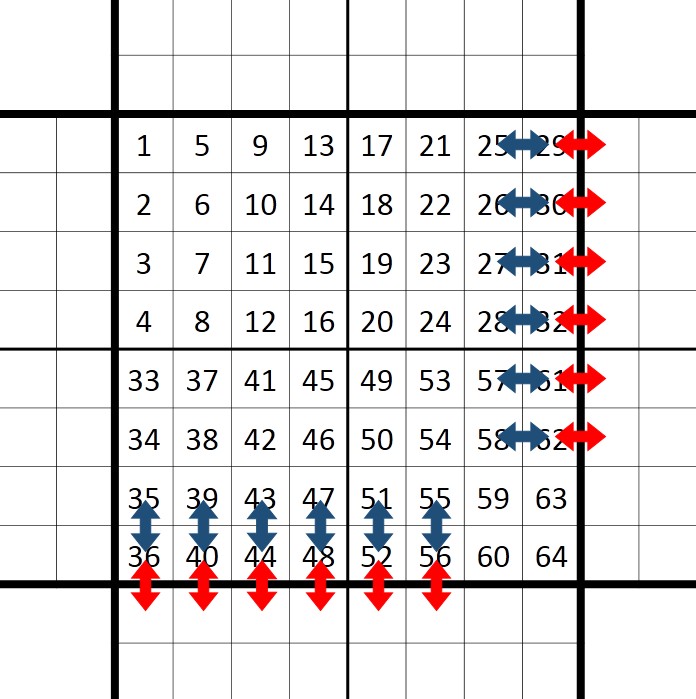}
\caption{Model of the video coding block with pixel numeration scheme}
\label{fig:moduleOftheVideoCodingBlock}
\end{figure}

Blocking is caused by independence of calculations for each block in 
the image. While many compression algorithms divide frames into blocks, this 
is one of the most popular and visible artifacts. Because of the coarse 
quantization, the correlation among blocks is lost, and horizontal and vertical 
borders appear. Another reason might be the resolution change, when a small 
picture is scaled up to be displayed on a larger screen.

Blockiness metric used in this work is based on \cite{Farias}. This metric assumes 
constant block size, which was chosen to be \num{8 x 8} pixels. Metric value depends on two factors: magnitude of color 
difference at the block’s boundary, and picture contrast near boundaries.

Consequently, $\mathit{InterSum}$ and $\mathit{IntraSum}$ values are computed for every in-coming frame.
\begin{enumerate}[(ii)]
    \item $\mathit{InterSum}$ is a sum of the absolute differences between pixels located 
on the border of two neighboring picture blocks, Eq.~(\ref{eq:intersum}).
    \item $\mathit{IntraSum}$ is a sum of the absolute differences between pixels located 
directly next to the neighboring pixel of the picture block, 
Eq.~(\ref{eq:intrasum}).
\end{enumerate}

\begin{center}
\begin{multline}
\mathit{InterSum}_{x,y}=\\
\left | b_{x,y}(29)-b_{x+1,y}(25) \right |+\left | b_{x,y}(30)-b_{x+1,y}(26) 
\right |+\left | b_{x,y}(31)-b_{x+1,y}(27) \right |+\\
\left | b_{x,y}(32)-b_{x+1,y}(28) \right |+\left | b_{x,y}(61)-b_{x+1,y}(57) 
\right |+\left | b_{x,y}(62)-b_{x+1,y}(58) \right |+\\
\left | b_{x,y}(36)-b_{x,y+1}(35) \right |+\left | b_{x,y}(40)-b_{x,y+1}(39) 
\right |+\left | b_{x,y}(44)-b_{x,y+1}(43) \right |+\\
\left | b_{x,y}(48)-b_{x,y+1}(47) \right |+\left | b_{x,y}(52)-b_{x,y+1}(51) 
\right |+\left | b_{x,y}(56)-b_{x,y+1}(55) \right |.
\label{eq:intersum}
\end{multline}

\begin{multline}
\mathit{IntraSum}_{x,y}=\\
\left | b_{x,y}(29)-b_{x+1,y}(1) \right |+\left | b_{x,y}(30)-b_{x+1,y}(2) 
\right |+\left | b_{x,y}(31)-b_{x+1,y}(3) \right |+\\
\left | b_{x,y}(32)-b_{x+1,y}(4) \right |+\left | b_{x,y}(61)-b_{x+1,y}(33) 
\right |+\left | b_{x,y}(62)-b_{x+1,y}(34) \right |+\\
\left | b_{x,y}(36)-b_{x,y+1}(1) \right |+\left | b_{x,y}(40)-b_{x,y+1}(5) 
\right |+\left | b_{x,y}(44)-b_{x,y+1}(9) \right |+\\
\left | b_{x,y}(48)-b_{x,y+1}(13) \right |+\left | b_{x,y}(52)-b_{x,y+1}(17) 
\right |+\left | b_{x,y}(56)-b_{x,y+1}(21) \right |.
\label{eq:intrasum}
\end{multline}
\end{center}

Computing scheme of $\mathit{InterSum}$ and $\mathit{IntraSum}$ is depicted in Fig. 
\ref{fig:moduleOftheVideoCodingBlock}, along with the pixel numeration scheme. $b_{x,y}(i)$ used in Eq.~(\ref{eq:intersum}) and (\ref{eq:intrasum}) means $i$-th pixel of $x,y$ block.
Blockiness metric is the ratio of $\mathit{IntraSum}$ to  $\mathit{InterSum}$, as presented by 
Eq.~(\ref{eq:blockiness}).

\begin{equation}
\mathit{blockinessMetric}=\frac{\sum_{y=2}^{\mathit{BLY}-1}\sum_{x=2}^{\mathit{BLX}-1}\mathit{IntraSum}_{x,y}}{\sum_{y=2}^
{\mathit{BLY}-1}\sum_{x=2}^{\mathit{BLX}-1}\mathit{InterSum}_{x,y}}.
\label{eq:blockiness}
\end{equation}

\subsection{Exposure time distortions}

\begin{figure}
\centering
\includegraphics[width=0.95\hsize]{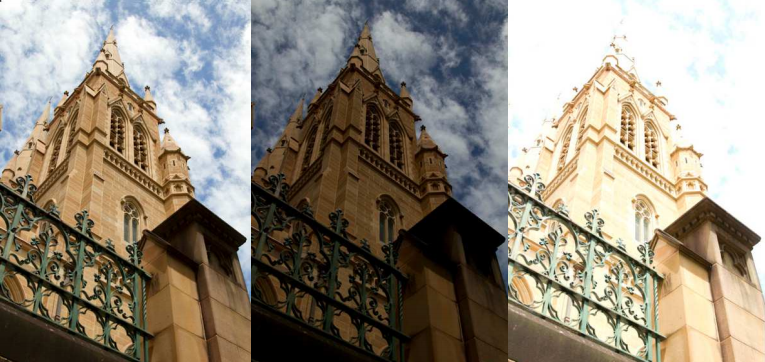}
\caption{Correct image(left), underexposed (center), overexposed (right)}
\label{fig:exposureLevels}
\end{figure}

\begin{figure}
\centering
\includegraphics[width=0.75\hsize]{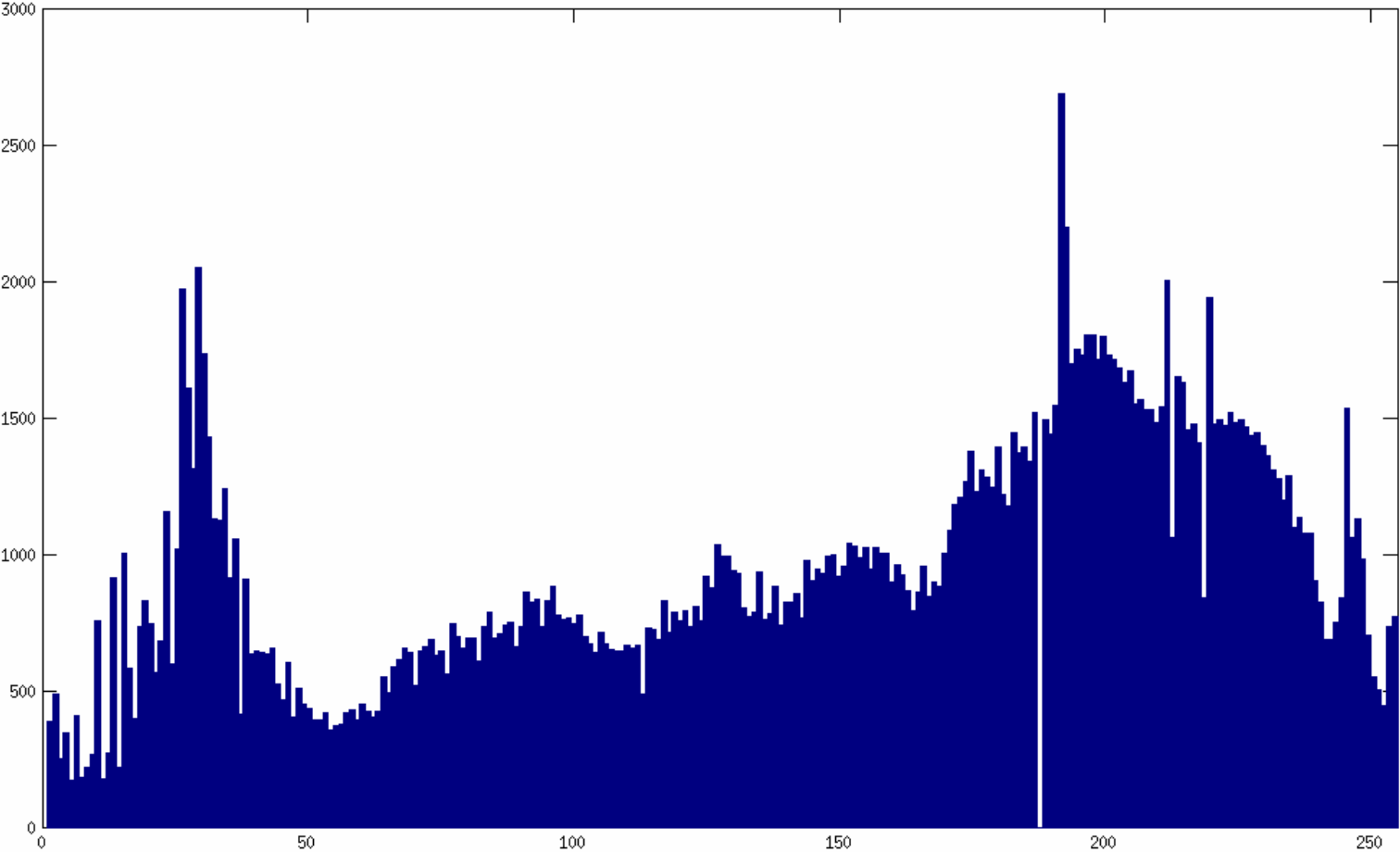}
\includegraphics[width=0.75\hsize]{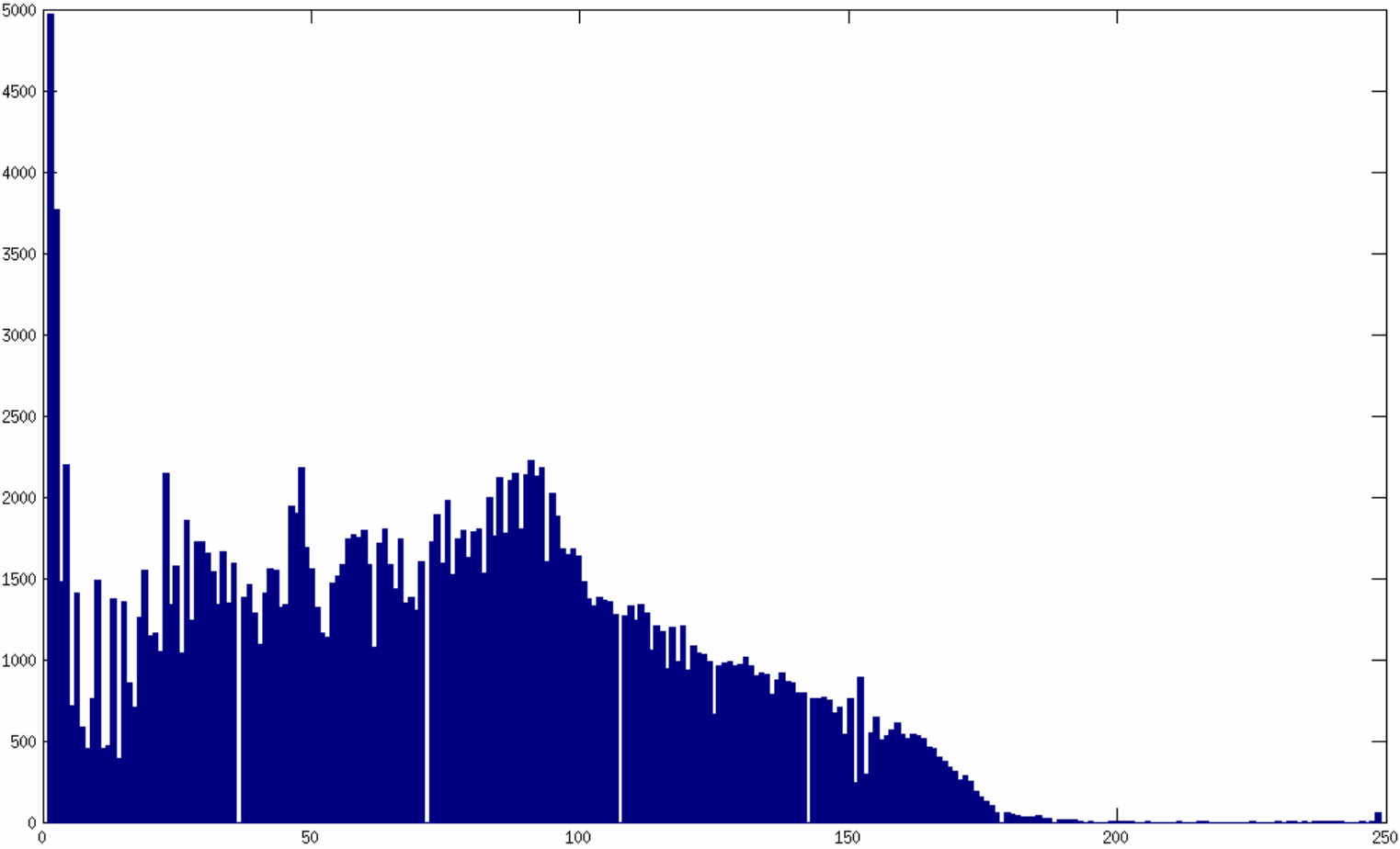}
\includegraphics[width=0.75\hsize]{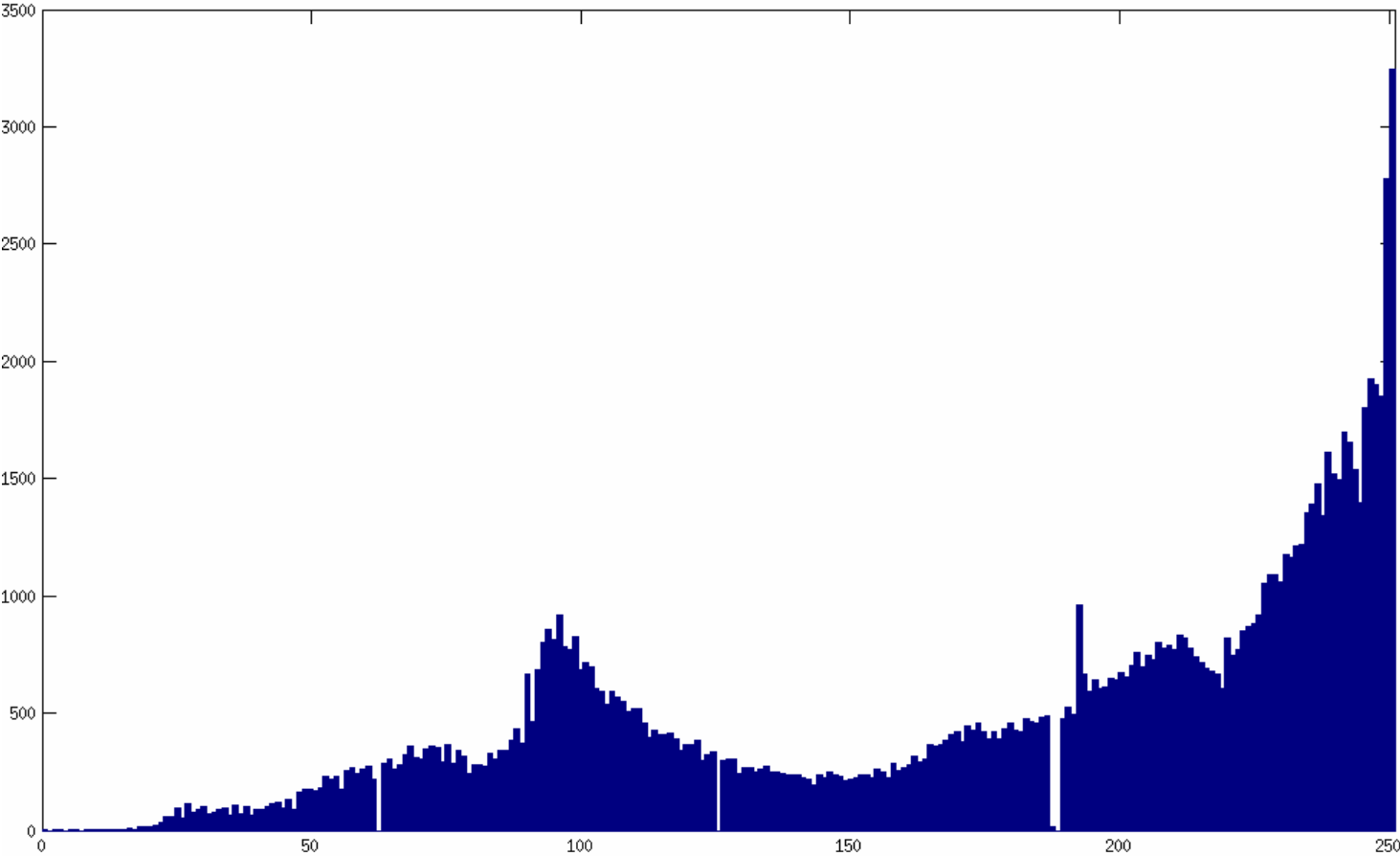}
\caption{Luminance histograms of correct (top), underexposed (middle) and overexposed (bottom) images}
\label{fig:exposure}
\end{figure}

Exposure time distortions are visible as an imbalance in brightness (frames 
that are too dark or too bright). They are caused by an incorrect exposure 
time, 
or recording a video without a sufficient lighting device. It is also possible to 
cause this distortion by improper digital enhancement. Various exposure levels 
for the same image are presented in Fig.~\ref{fig:exposureLevels}. Histograms 
of luminance for each of those images are presented in Fig.
\ref{fig:exposure}.

Mean brightness of the darkest and brightest parts of the image is calculated 
in order to detect the distortion. Exposure metric is presented in Eq. 
(\ref{eq:exposureMetric}), where L$_{d}$, Eq.~(\ref{eq:exposureDark}), 
represents three darkest blocks, L$_{b}$, Eq.~(\ref{eq:exposureBright}), 
represents three brightest blocks.

\begin{equation}
 \mathit{exposureMetric}=\frac{L_{b}+L_{d}}{2}.
 \label{eq:exposureMetric}
\end{equation}

\begin{equation}
 L_{d}=\sum_{i=1}^{3}\mathit{sortMeanBL}_{i}.
 \label{eq:exposureDark}
\end{equation}

\begin{equation}
 L_{b}=\sum_{i=\mathit{BLX}\times \mathit{BLY}-2}^{\mathit{BLX}\times \mathit{BLY}}\mathit{sortMeanBL}_{i}.
 \label{eq:exposureBright}
\end{equation}

\glsreset{MOS}
The results of the metrics mentioned above 
were mapped to the \gls{MOS}. The thresholds were referred to 
the \gls{MOS} scale, determining the score below which each distortion is noticeable.

\subsection{Blackout}

It is manifested as the picture disappearing; a black screen. It appears when 
all packets of data are lost, or as a result of incorrect video recording.
Image blackout detection is independent of the frame color, i.e. detection result 
is positive (equals `1') if the frame has a uniform color, otherwise the result 
is `0'. Comparison of all the pixels of the frame under consideration seems to 
be the most straightforward approach. However, this is the greedy method which 
requires \emph{n} comparisons, where \emph{n} is the number of pixels within 
the frame. The authors came up with an alternative method which utilizes 
partial results of the exposure time distortion method.
This results in a significant reduction of the metric implementation cost.

The novel metric description:

A frame is split into blocks of \num{8 x 8} pixels. Sum of the luminance is 
calculated for every block. If the difference between the block of the highest 
luminance and the lowest is lower than the $\mathit{thBlout}$ threshold the 
detection result equals `1', otherwise it is `0'; $\mathit{thBlout}$ is set to a constant 
four.

\begin{equation}
\mathit{blackoutMetric}=\begin{cases}
0 \text{~if~} \mathit{sortSumBL}_{\mathit{BLX}\times \mathit{BLY}}-\mathit{sortSumBL}_{1}\geqslant \mathit{thBlout}\\
1 \text{~otherwise}
\end{cases}.
 \label{eq:blackout}
\end{equation}

\subsection{Interlace}

\begin{figure}
\centering
\includegraphics[width=4.0in]{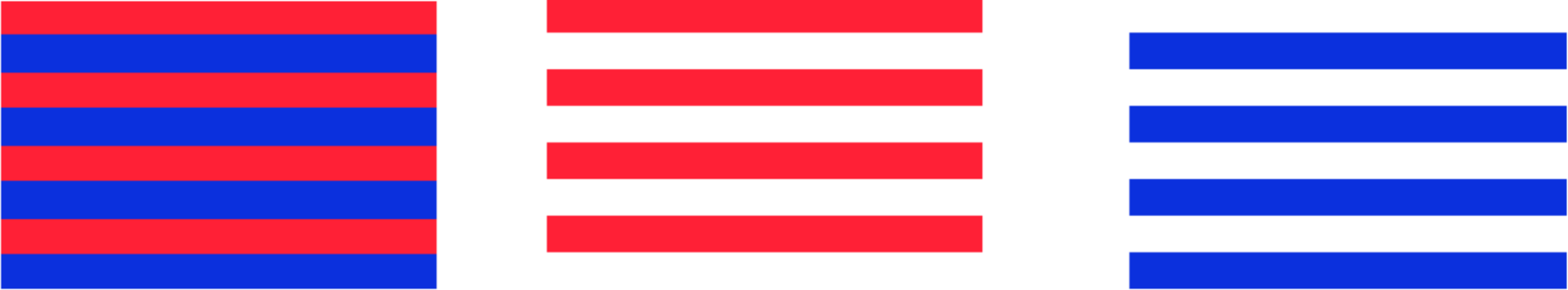}
\caption{Creation of interlaced frame}
\label{fig:interlaceExample}
\end{figure}

\begin{figure}
\centering
\includegraphics[width=1.0in]{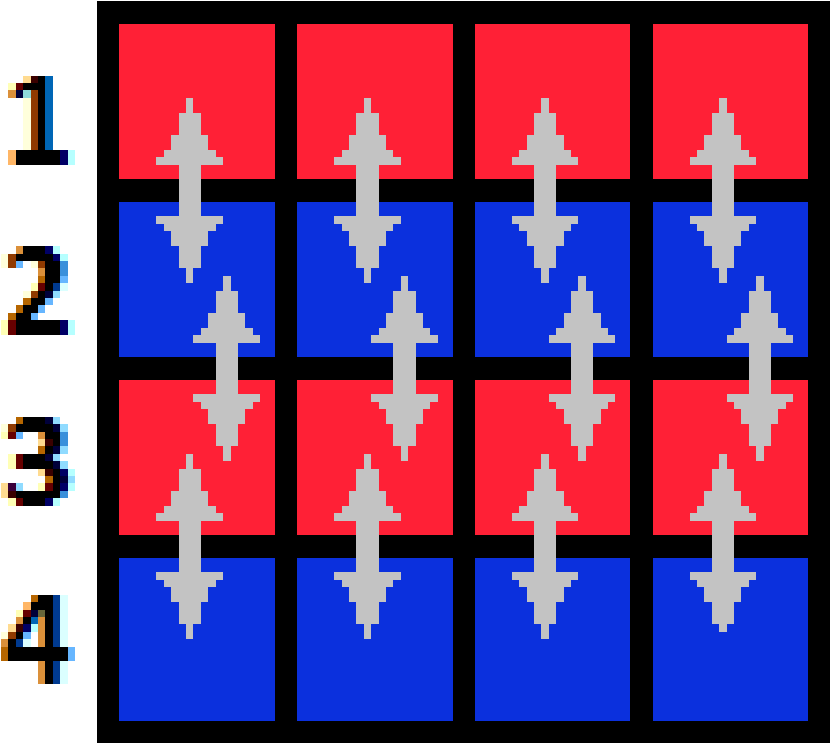}
\caption{Interlaced microblock model}
\label{fig:interlaceModel}
\end{figure}

Interlace is a technique where a single frame is a composition of two 
half-frames, each of which contains half of the information. Odd half-frames 
contain odd rows of pixels, while even half-frames contain even rows of pixels. 
Resulting frame is created by interlacing both of them. The idea of 
interlace is presented in Fig.~\ref{fig:interlaceExample}. Interlace 
distortion becomes visible when two half-frames are not properly 
aligned. It is especially visible for videos including motion.

\begin{figure}
\centering
\includegraphics[scale=0.57]{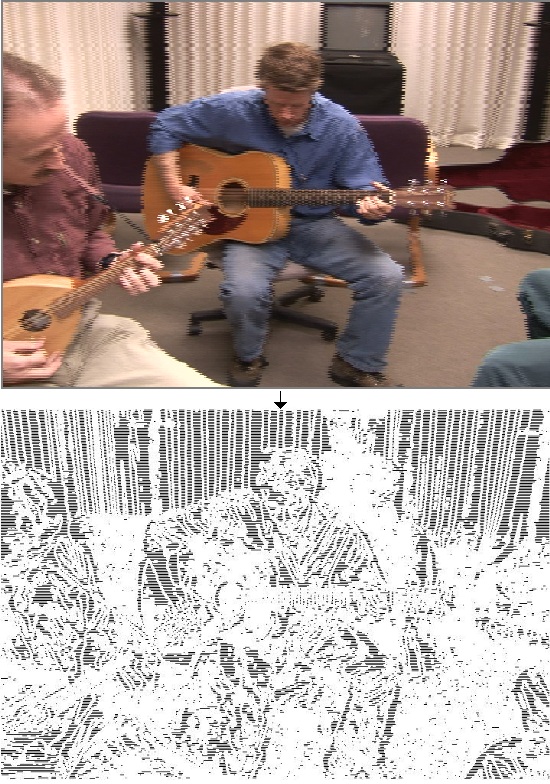} 
\caption{Detection of interlace artifact}
\label{fig:InterFig}
\end{figure}

The authors proposed their own solution for interlace distortion metric. It is calculated 
independently for each micro \num{4 x 4} pixel block and then subsequently combined 
into a complete metric. Given block is marked as a block with interlace distortion if 
change of luminance of the first row relative to the second row is in the same 
direction for all pixel pairs, change between second and third is in opposite 
direction, and change between third and fourth is in the same direction. All 
comparisons are presented in Fig.~\ref{fig:interlaceModel}.

\begin{equation}
\mathit{interlace}_{i}=\begin{cases} 
1 \text{~if~} \sum_{j=1}^{12}\left | \mathit{sgn}\left ( d_{i,j} \right ) \right |=12\\ 
0 \text{~otherwise}
\end{cases}.
\label{eq:interlace}
\end{equation}

\begin{equation}
\mathit{interlaceMetric} =\frac{\sum_{i=1}^{4\times \mathit{BLX}\times \mathit{BLY}}intelace_{i}}{4\times 
\mathit{BLX}\times \mathit{BLY}}.
\label{eq:intelaceMetric}
\end{equation}

Eq.~(\ref{eq:interlace}) determines if a given block has interlace 
distortion, where $d_{i,j}$ is the $j$-th difference between luminance values of the $i$-th micro block. Eq.~(\ref{eq:intelaceMetric}) calculates metric value for the 
whole frame.  Fig.~\ref{fig:InterFig} illustrates detection of interlace in a 
sample frame. The effect is the most visible in shapes containing sharp vertical lines. 
Presented solution shows positive results.

\section{High level hardware design -- tools and methodology}

The module was implemented using Impulse~C language. Impulse~C is a high level language based on Stream-C compiler, which was 
created in the Los Alamos National Laboratories in the 1990s. 
The idea  evolved into a corporation named Impulse Accelerated Technologies 
Company (2002), which is now a supporting vendor of Impulse 
C and holder of the Impulse~C rights. The main intention of the language 
designers was to bridge the gap between hardware and software 
and facilitate the process of system level design. It was achieved through 
abstracting most of the language constructs, so the designer 
can focus on the algorithm, rather than low level details of the implementation 
\cite{Offloading}.

There is a whole set of high level languages such as Dime-C, SystemC, Handel-C, 
Mitrion~C available nowadays, 
which enable specification and implementation of the system at the module 
level. However, most of them introduce their own structures 
(e.g. Mitrion~C), expanding or modifying existing standards of high level 
languages. On the one hand, such an approach 
helps to establish a design space by imposing a strict language expression set. 
On the other hand, designers have to comprehend a whole 
range of the language structures, along with their appropriate application 
schemes, which may be pretty tedious. Such an extra effort 
is justified in the case of people, who expect to use the tool for a reasonably 
long time (professional digital logic designers). 
Unfortunately, most of the \gls{FPGA} \gls{HLL} users are people familiar with programming 
languages (e.g. C, C++, Java, Fortran), who need to 
port some part of their application into hardware. Therefore, it seems 
reasonable to 
leverage one of the well-known standards such as ANSI~C. Moreover, ANSI~C allows 
for access to low level details of an application,
which is very useful in some cases. It can be said, that \emph{C gives the 
lowest possible level of abstraction among the high 
level languages}. The aforementioned ideas prevailed in the design of Impulse~C.

There are several features of the Impulse~C language, which in the authors' 
view 
are superior to other currently used \glspl{HLL}. First of
all, Impulse~C allows designers to reuse their HDL code by providing 
mechanisms, 
which facilitate the incorporation of existing modules.
Furthermore, three different architectures are supported: 
\emph{combinational}, \emph{pipelined} and \emph{asynchronous}, which
cover a complete range of existing design scenarios. Secondly, C compatibility 
makes it easy for software engineers to switch from \gls{GPP} programming to \gls{FPGA} design, as well as 
providing a platform for software-hardware integration
within one design environment. Finally, Impulse~C comes with a range of \glspl{PSP}, which provide a
communication interface between \gls{FPGA} and \gls{GPP} computational nodes. Furthermore, 
\glspl{PSP} usage provides portability of an application across
different platforms. In fact, \glspl{PSP} are packs of files that describe a system's 
profile 
to the Impulse~C compiler \cite{PSP}. The compiler uses this information to 
generate interface components needed to connect hardware
processes to a system bus and interconnect them together inside the \gls{FPGA}, and 
also to establish the software side of any software/hardware
stream, signal, memory, etc. connections \cite{PSP,Papu09,ImpC}.

The language enables both fine-grain and course-grain parallelism. The former 
one is implemented within a process, whereas the latter
is built as multiple-process structures.

It is worth noting, that algorithm partitioning must be handled by a programmer: this stage is not automated by the compiler,
which means that it is up to the designer to classify different sections of an 
application. However, due to portability of a code,
it is possible to migrate between hardware and software sections, if adequate 
language structures are employed. With respect to
this, it is recommended to avoid using language constructs which confine a 
given part of the code to software or hardware solely.

A designer should keep a number of control signals and branches low,
since the primary goal of \gls{HLL} \gls{FPGA} algorithm implementation is
to increase throughput at the expense of latency (trade latency for 
throughput). Using control signals may compromise this effort
and should make a designer rethink a concept for the architecture.

Impulse~C compiler automatically generates test benches, software-hardware 
interfaces and synthesisable HDL code; it automatically
finds parallel structures in the code as well. However, it is good coding 
practice to explicitly point out sections which are to
be paralleled. Both hardware and software parts of the code can be compiled 
with \gls{GCC}.

\begin{figure}
\centering
\includegraphics[width=0.8\hsize]{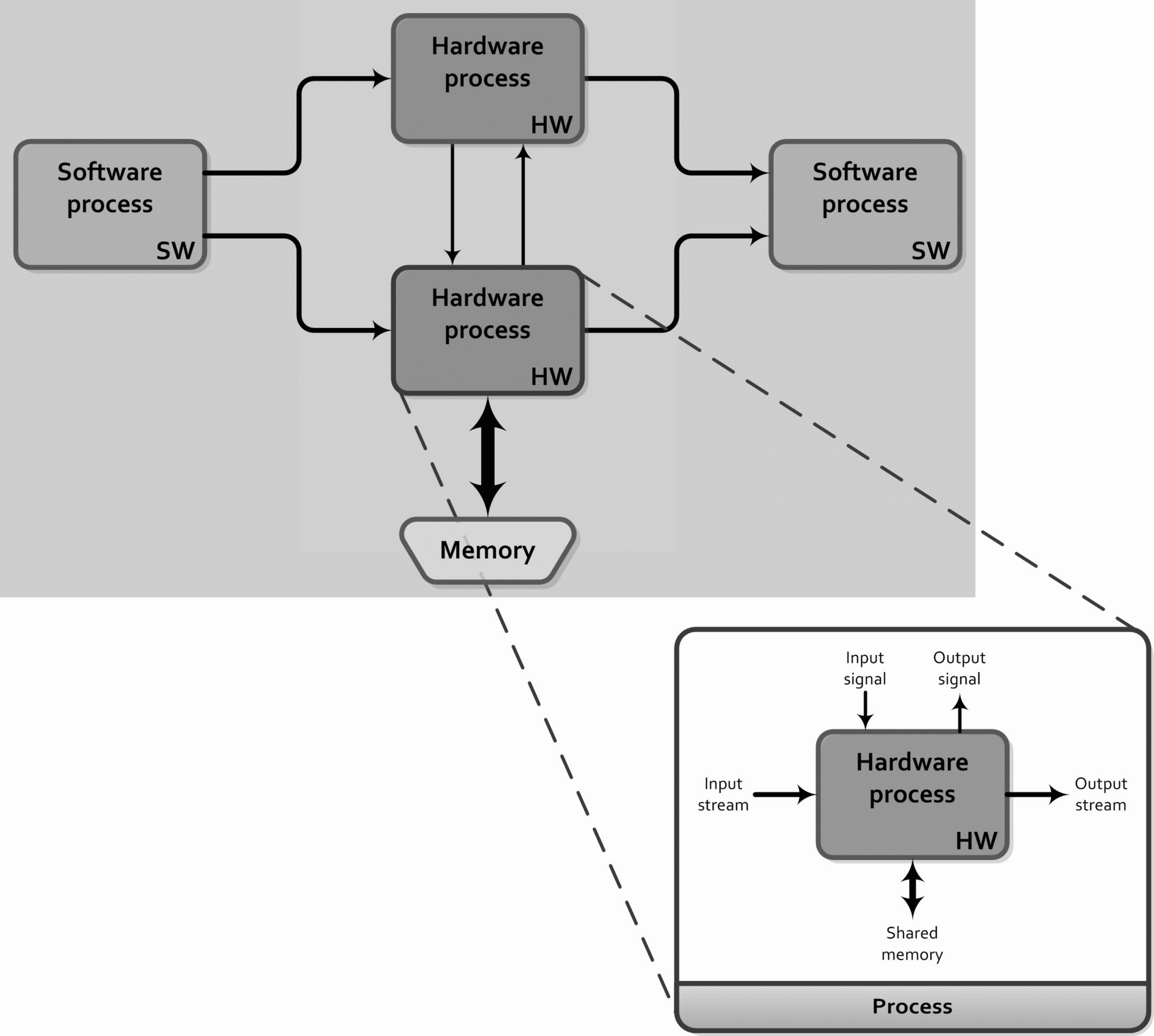}
\caption{Impulse~C programming model.}
\label{fig:ImpCMod}
\end{figure}

Impulse~C can be characterized as a stream-oriented, process based language. 
Processes are main building blocks interconnected
using streams to form an architecture for the desired hardware module. From the 
hardware perspective, processes and streams are
hardware modules and \gls{FIFO} registers, 
respectively. 
The Impulse~C programming model is based on 
the \textit{Communicating Sequential Processes} model\cite{ImpC} and is 
illustrated in Fig.~\ref{fig:ImpCMod}. Every process
must be classified as a hardware or a software process.

It is the programmer's responsibility to ensure the interprocess 
synchronisation. Like most of the \glspl{HLL}, Impulse~C does not provide
access to the clock signal, which relieves the designer from implementing cycle 
synchronization procedures. However, it is possible 
to attach HDL modules and synchronize them at the level of RTL using clock 
signal.

\section{FPGA-based platform}

\begin{figure}
\centering
\includegraphics[width=4.0in]{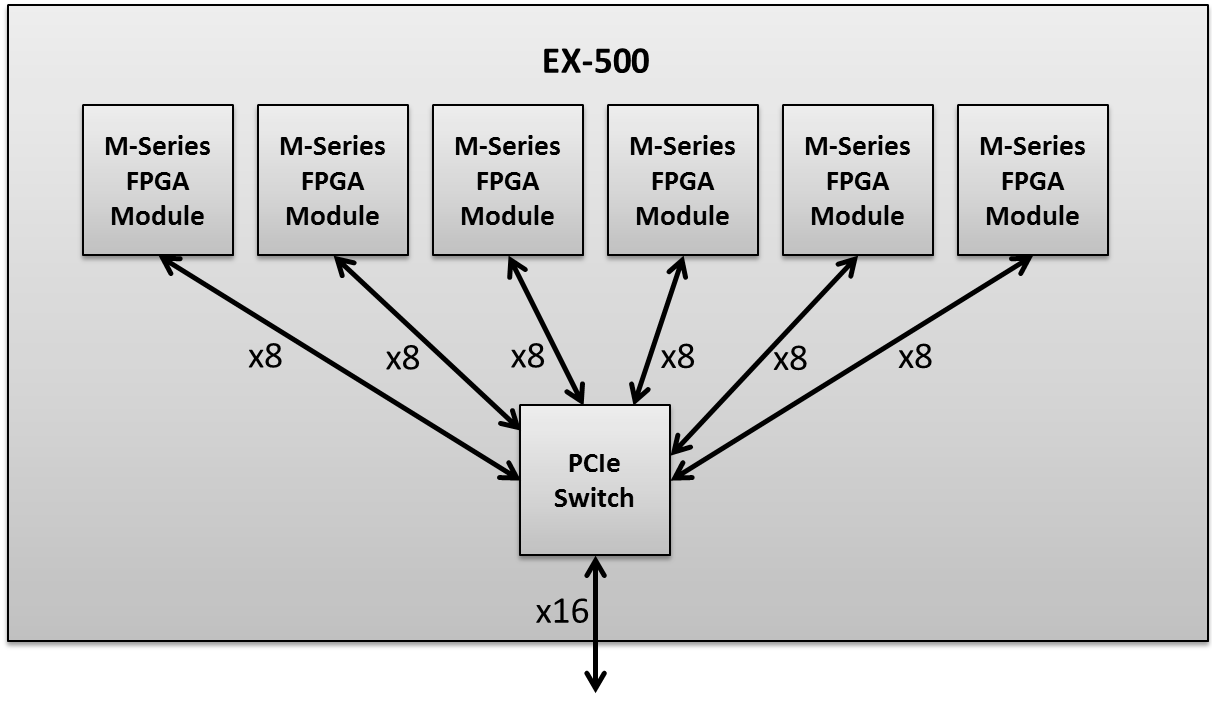}
\caption{The FPGA-based platform used for the computations -- Pico Computing}
\label{fig:PicoFig}
\end{figure}

The module was implemented on Pico M503 platform\cite{Pico}, connected 
through PCIe to server with Intel i7-950 processor and \SI{12}{\giga\byte} of RAM. The Pico 
platform (Fig.~\ref{fig:PicoFig}) consists of two components:
\begin{enumerate}[(ii)]
\item EX-500 board with a Gen2 PCI-Express controller which enables connecting 
up to six \glspl{FPGA} to the motherboard
\item M503 FPGA boards\cite{Pico}
\end{enumerate}

Communication between a CPU and the \gls{FPGA} is realized with eight lines of PCIe 
interface –- full-duplex connection streams. 
If more than two boards are used, the throughput is limited to \SI[per-mode=symbol]{5}{\giga\byte\per\second}; 
in case of using only one board the maximum throughput reaches about \SI[per-mode=symbol]{3}{\giga\byte\per\second}. 
Another limitation is the width of the stream, which is equal to \num{128} bits.

\section{Impulse~C implementation of the module}
This section is a description of implementation of hardware version of the video quality module. Subsection \ref{sub:generalPart}  shows a general concept of the module, next the description is divided into two parts according to two parts of projects in Impulse~C language: software (\ref{sub:softPart}) and hardware (\ref{sub:hardPart}).

\subsection{Architecture of the module}
\label{sub:generalPart}

\begin{figure}
\centering
\includegraphics[width=4.0in]{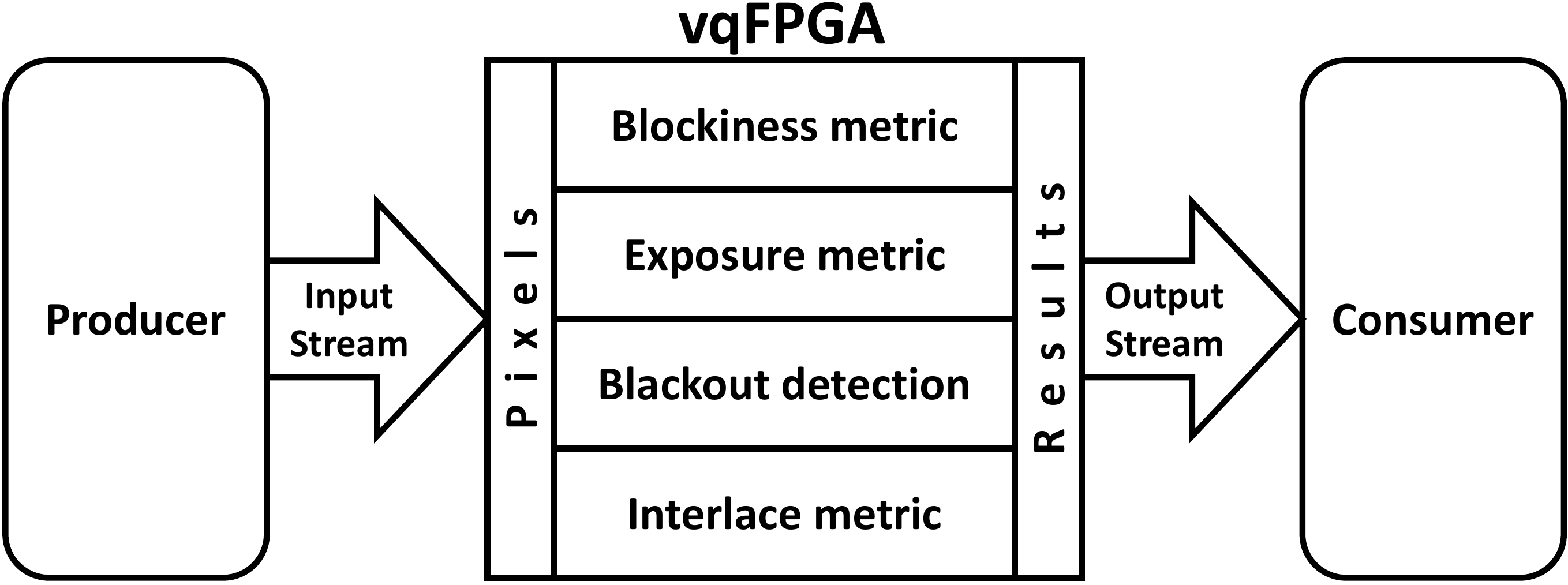}
\caption{Architecture of the video quality assessment module as implemented in 
Impulse~C}
\label{fig:general}
\end{figure}

The block diagram of the video quality assessment module is shown in Fig. 
\ref{fig:general}. It consists of three subblocks:

\begin{enumerate}[(ii)]
\item	Producer -– reads video data from a file and sends it to the vqFPGA 
block using the InputStream
\item	vqFPGA –- reads data from the InputStream, executes video quality 
metrics and sends 
results to the Consumer process using the OutputStream
\item Consumer –- reads data from the OutputStream, analyses it and sends them 
to standard output stream.
\end{enumerate}

Width of the Input and the Output stream is \num{128} bits, which is the maximum 
width of Pico M503 platform stream. The scheme described above is parallelized 
sixfold, in the real module there are six producer, vqFPGA and consumer 
processes.

Every Impulse~C project is composed of a software and hardware part, 
and so is the video quality assessment 
module.

\subsection{Software part of the module}
\label{sub:softPart}

\begin{figure}
\centering
\includegraphics[width=3.5in]{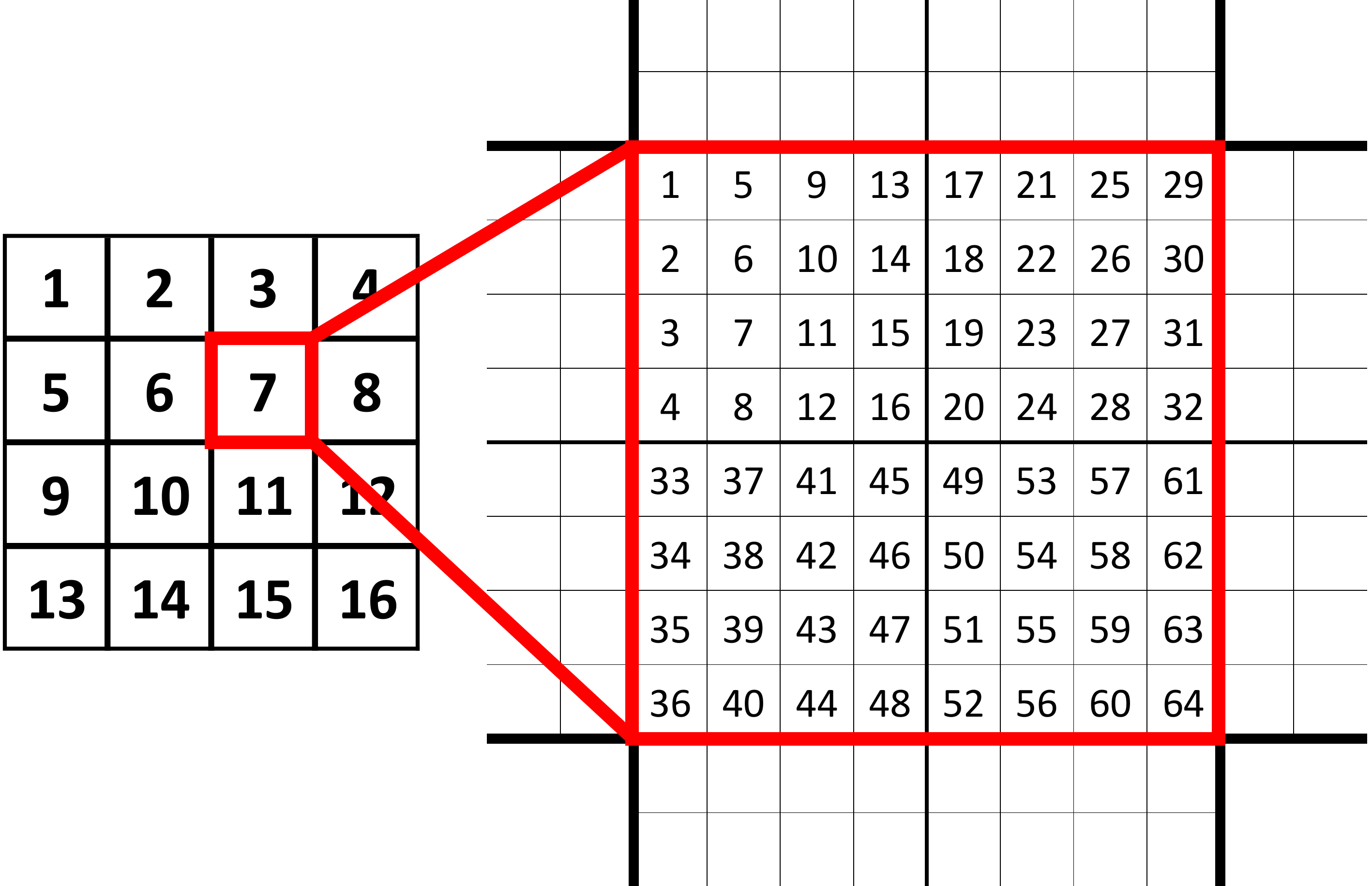}
\caption{A structure of a sample block and corresponding transfer sequence 
of a sample frame (left) and an order of pixels within each block (right)}
\label{fig:structureOfsample}
\end{figure}

Software part is composed of three functions: producer, consumer and the main 
program function which is used to launch the FPGA-based accelerator and all the 
application-related threads. It is also responsible for programming the FPGA 
with a bit file. Producer function opens input stream to FPGA and sends pre 
read video data. Pico module input stream is \num{128} bit wide, thus it is 
recommended to organize the data in such chunks so the best possible throughput 
is achieved.
Every \num{8 x 8} block is divided into four microblocks. Every microblock 
contains \num{16} values, eight bits each. Such structure allows for sending whole 
block in four bus clock cycles, retaining data consistency. Described scheme is 
presented in Fig.~\ref{fig:structureOfsample}.

\begin{figure}
\centering
\includegraphics[width=3.5in]{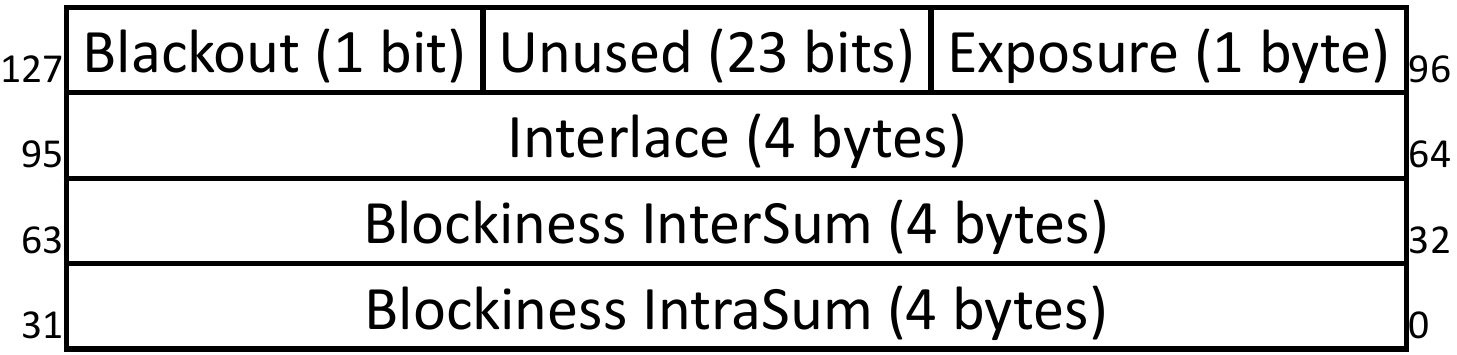}
\caption{Structure of the frame sent through OutputStream}
\label{fig:outputStream}
\end{figure}

Consumer function manages module output stream. At the end of every video frame, 
a valid results frame is received. Its size is also fixed to \num{128} bits wide, as 
it fits best to hardware. A special structure of the results frame was 
designed, as presented on Fig.~\ref{fig:outputStream}. The frame contains the 
results of calculation of blackout, exposure and interlace distortion metrics. The last 
part of computing the blockiness metric is performed in software, thus frame 
contains required $\mathit{InterSum}$ and $\mathit{IntraSum}$.

\subsection{Hardware part of the module}
\label{sub:hardPart}
Hardware part is composed of vqFPGA modules and the additional hardware which 
handles data fetching and sending results to the software part. Hardware part 
is equipped with two data streams corresponding to software streams, which are 
opened before the data transfer is conducted and closed once it is finished. 
Hardware module requires the information about video resolution to be sent 
in advance to the actual stream. Every \num{128} bit word is then arranged into a
microblock. Afterwards, data is sent to the parts of hardware 
responsible for computing each metric.
The module registers are reset after all the microblocks of a given frame are 
processed and a new frame comes in. The maximum number of combinational stages 
between registers were experimentally determined as \num{64} and implemented with 
\texttt{Co Set stageDelay} Impulse~C pragma. This also requires using 
\texttt{Co 
Pipeline} pragma which implements pipelined design approach.

\subsubsection{Blockiness metric}
\label{sub:blockiness}

\begin{figure}
\centering
\includegraphics[width=3.0in]{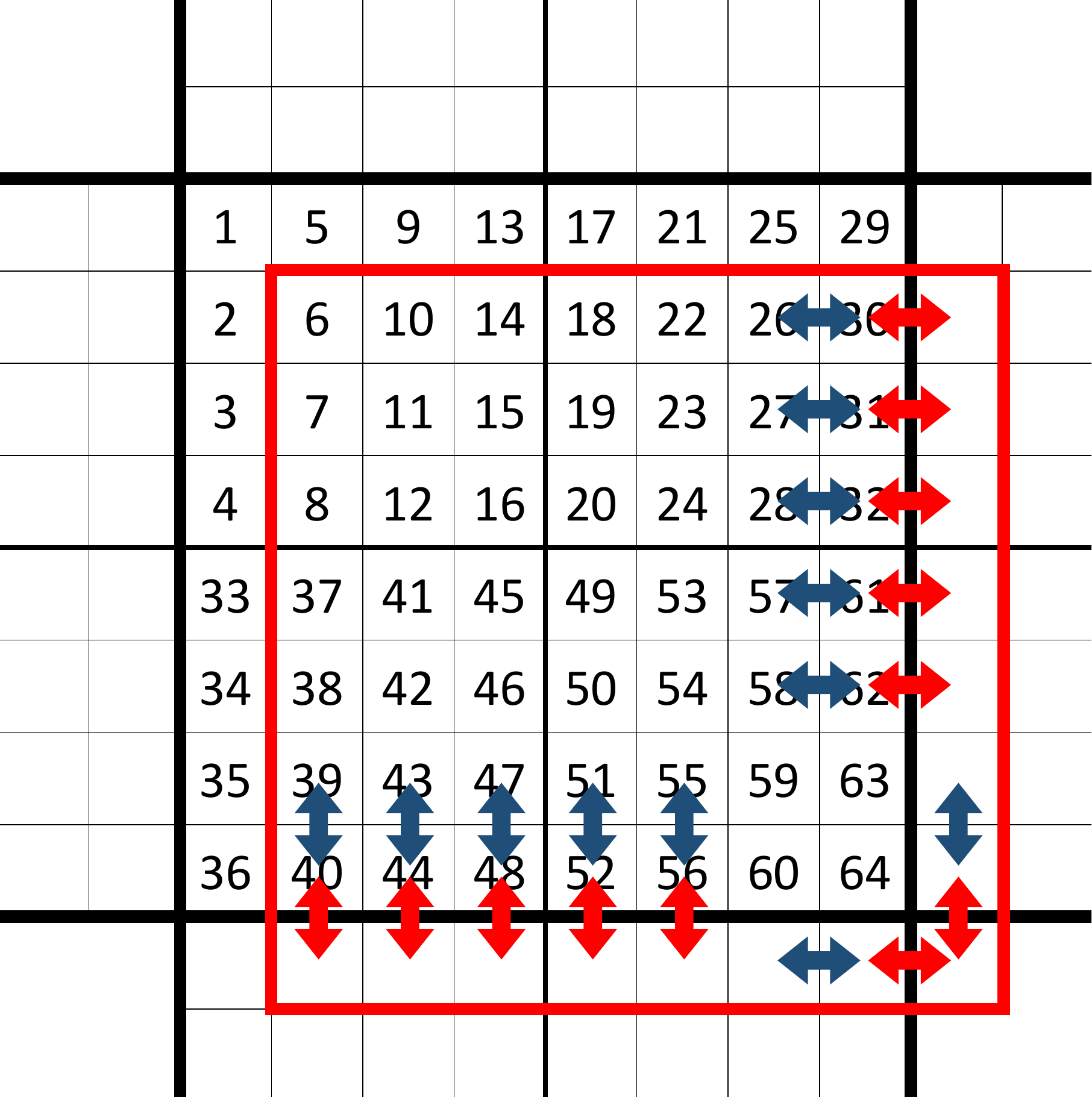}
\caption{Shifted block}
\label{fig:shiftedblock}
\end{figure}

For blockiness metric, only the most computationally demanding parts were 
implemented in hardware. $\mathit{InterSum}$ and $\mathit{IntraSum}$ are calculated inside \gls{FPGA} while 
final division is done in software. As presented in Fig. 
\ref{fig:moduleOftheVideoCodingBlock}, calculations require data from 
neighboring blocks and storing all necessary data inside \gls{FPGA} would be very 
inconvenient. Therefore, authors modified data sending scheme to make it more 
suitable for blockiness metric calculation. First row and first column are 
omitted and block boundaries are shifted as presented in Fig.~\ref{fig:shiftedblock}. After such operations, all data necessary for $\mathit{InterSum}$ and $\mathit{IntraSum}$ calculations are available in a single block.

The source code presented in Fig. \ref{fig:blockinesCode} shows hardware 
implementation of 
the blockiness metric. Due to the efficient data serialization, the module is 
implemented with few lines of code, which also results in low hardware 
resources consumption. It is worth noting that the source code reflects the 
operations described by Eq.~(\ref{eq:blockiness}).

\begin{figure}
\begin{lstlisting}
if(microBlock % 4 == 2){
  IntraSum += ABSDIFF(p9,p5) + ABSDIFF(p10,p6) + ABSDIFF(p11,p7) + ABSDIFF(p12,p8);
  InterSum += ABSDIFF(p9,p13) + ABSDIFF(p10,p14) + ABSDIFF(p11,p15) + ABSDIFF(p12,p16);
}
if(microBlock % 4 == 3){
  IntraSum += ABSDIFF(p2,p3) + ABSDIFF(p6,p7) + ABSDIFF(p10,p11) + ABSDIFF(p14,p15);
  InterSum += ABSDIFF(p4,p3) + ABSDIFF(p8,p7) + ABSDIFF(p12,p11) + ABSDIFF(p16,p15);
}
if(microBlock % 4 == 0){
  IntraSum += ABSDIFF(p9,p5) + ABSDIFF(p8,p12) + ABSDIFF(p2,p3) + ABSDIFF(p14,p15);
  InterSum += ABSDIFF(p9,p13) + ABSDIFF(p12,p16) + ABSDIFF(p4,p3) + ABSDIFF(p15,p16);
}
\end{lstlisting}
\caption{Blockiness metric source code}
\label{fig:blockinesCode}
\end{figure}

\subsubsection{Exposure time distortions metric}
\label{sub:exposure}
The metric is composed of three steps. In the first one, a luminance mean value 
of every code block is calculated. Then, six extreme 
values for every frame are found (three smallest and three biggest). 
The extreme values are used to compute the mean value.

Several modifications were introduced to adapt it to hardware implementation. A 
size of each block is constant, therefore instead of the mean, a sum of values 
may be used. This allows to eliminate the division operation in mean 
calculation which is very resource demanding. It is only performed for the 
border blocks. Fractional part may be disregarded as of little importance. 
Without changing algorithm, the mean may be computed for eight 
results (four biggest and four smallest). This will enable the use of bit 
shift (shift right by two bits) operation instead of very hardware expensive 
division.

Sum of luminance values is stored in a $\mathit{blockSum}$ variable. The extreme 
blocks are searched for (Fig.~\ref{fig:exposure1}) and the sum of their luminance 
values are stored in the following variables $\mathit{blockSumMAX1}$, 
$\mathit{blockSumMAX2}$, $\mathit{blockSumMAX3}$, $\mathit{blockSumMAX4}$ and $\mathit{blockSumMIN1}$, 
$\mathit{blockSumMIN2}$, $\mathit{blockSumMIN3}$, $\mathit{blockSumMIN4}$.

\begin{figure}
\begin{lstlisting}
if(blockSum < blockSumMIN4){
    if (blockSum < blockSumMIN3){
        if(blockSum < blockSumMIN2){
            if(blockSum < blockSumMIN1){
                blockSumMIN4 = blockSumMIN3;
                blockSumMIN3 = blockSumMIN2;
                blockSumMIN2 = blockSumMIN1;
                blockSumMIN1 = blockSum;
            }
            else {
                blockSumMIN4 = blockSumMIN3;
                blockSumMIN3 = blockSumMIN2;
                blockSumMIN2 = blockSum;
            }
        }
        else {
            blockSumMIN4 = blockSumMIN3;
            blockSumMIN3 = blockSum;
        }
    }
    else
        blockSumMIN4 = blockSum;
}
\end{lstlisting}
\caption{Implementation of the exposure metric - minimal luminance values of the frame}
\label{fig:exposure1}
\end{figure}

Result of the metric is a weight mean of the pixels luminance from extreme 
blocks, i.e. all the $\mathit{blockSumMAX}$ and $\mathit{blockSumMIN}$ are summed up
and the result is shifted left by nine bits (nine because $2^9 = 512 = 8*64$; 
eight is a number of extreme blocks and \num{64} is a number of pixels within a 
single block). In order to prevent data range overflow (\texttt{co\_uint16} is used) 
each datum is shifted right by two bits and the result is subsequently moved 
by the remaining seven bits. $\mathit{microBlock}$ and $\mathit{blockSumMAX}$ are reset after all the data 
results are sent to the software part of the module.
$\mathit{blockSumMIN}$ is set to \num{16384} before the next frame is taken from the input.

\subsubsection{Blackout metric}
\label{sub:blackout}

Blackout metric is implemented as four lines of Impulse~C code (Fig. 
\ref{fig:blackout1}). The module comprises one adder/subtractor and one 
comparator. The metric result is sent to the software part of the module as a 
single bit set to `1' in OutputStream which indicates that 
blackout occurred.  

\begin{figure}
\begin{lstlisting}
if((blockSumMAX1 - blockSumMIN1) > thBlout)
    blackoutMetric = 0;
else
    blackoutMetric = 1;
\end{lstlisting}
\caption{Implementation of blackout metric}
\label{fig:blackout1}
\end{figure}

\subsubsection{Interlace distortion metric}
\label{sub:Interlace}

The way data is structured and transferred between hardware and software part 
is 
presented in subsection \ref{sub:softPart}.
It is adapted to this particular metric and improves performance of the module. 
A single microblock is sent and the interlace distortion detection is conducted just by examining $\mathit{IS\_INTERLACE}$ and $\mathit{IS\_INTERLACE2}$ (Fig.~\ref{fig:interlace1}) conditions.

\begin{figure}
\begin{lstlisting}
#define IS_INTERLACE (((p1>p2) && (p5>p6) && (p9>p10) && (p13>p14)\
            && (p3>p2) && (p7>p6) && (p11>p10) && (p15>p14)\
            && (p3>p4) && (p7>p8) && (p11>p12) && (p15>p16)))
                                 
#define IS_INTERLACE2 ((p1<p2) && (p5<p6) && (p9<p10) && (p13<p14)\
            && (p3<p2) && (p7<p6) && (p11<p10) && (p15<p14)\
            && (p3<p4) && (p7<p8) && (p11<p12) && (p15<p16))
\end{lstlisting}
\caption{Implementation of interlace distortion metric}
\label{fig:interlace1}
\end{figure}

If one of those conditions is met, the result of the metric is incremented by 
one. 
Sum of all the microblocks of a frame is a max.
possible value of the result which affected a choice of the variable used to 
store it, i.e. \texttt{co\_uint32}. After all the microblocks 
of the frame are received, the variable is reset. The module is composed of \num{12} 
interconnected comparators which form a single 
huge XNOR gate. In addition, the module comprises an adder and \num{32}-bit shift 
register for $\mathit{interlaceMetric}$ variable. 

\section{Experimental results}

\begin{figure}
\centering
\includegraphics[width=\hsize]{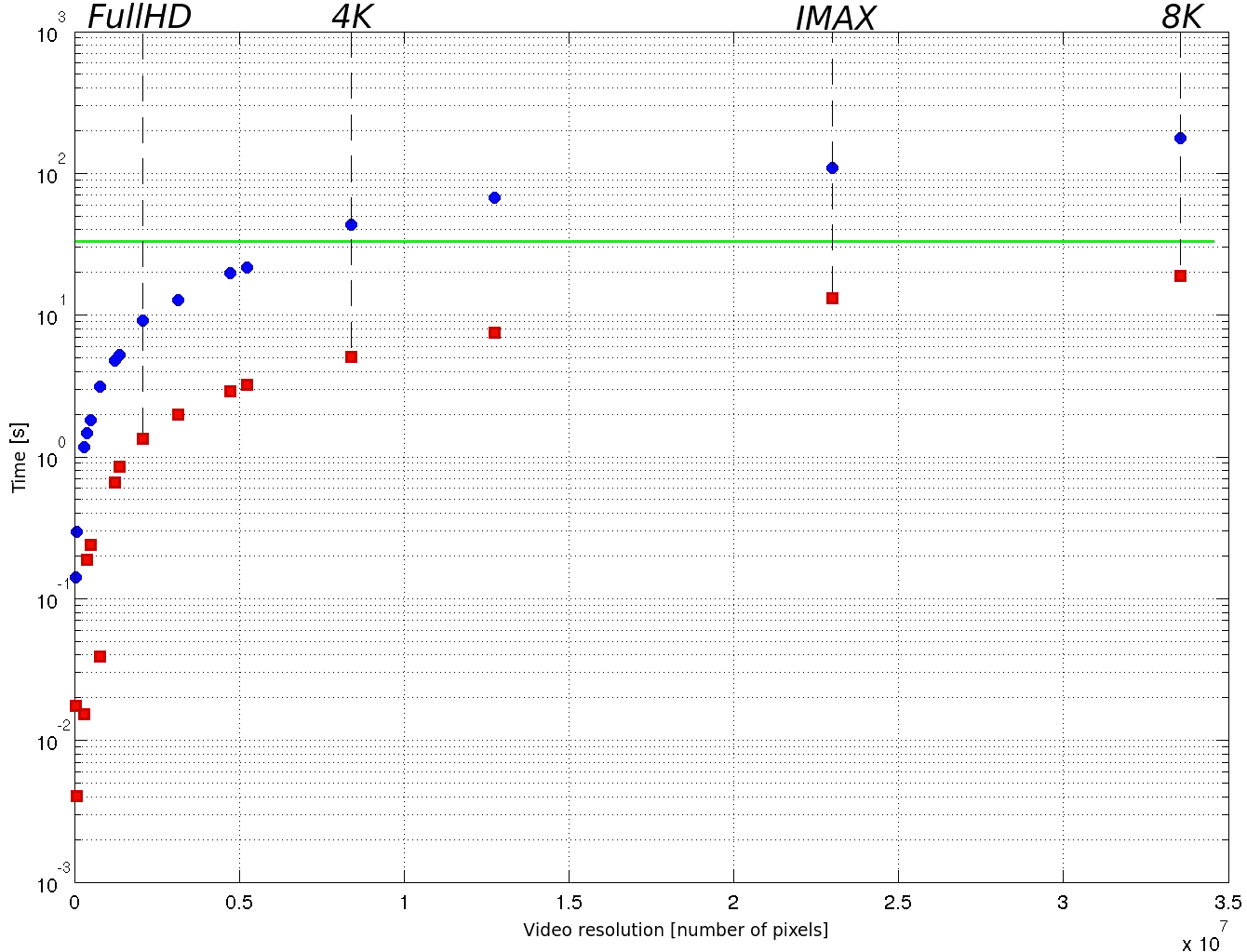}
\caption{Processing time of \num{1000} frames as a function of video resolution for 
both hardware (red squares) and software (blue dots) implementations}
\label{fig:result1}
\end{figure}

\begin{figure}
\centering
\includegraphics[width=5.0in]{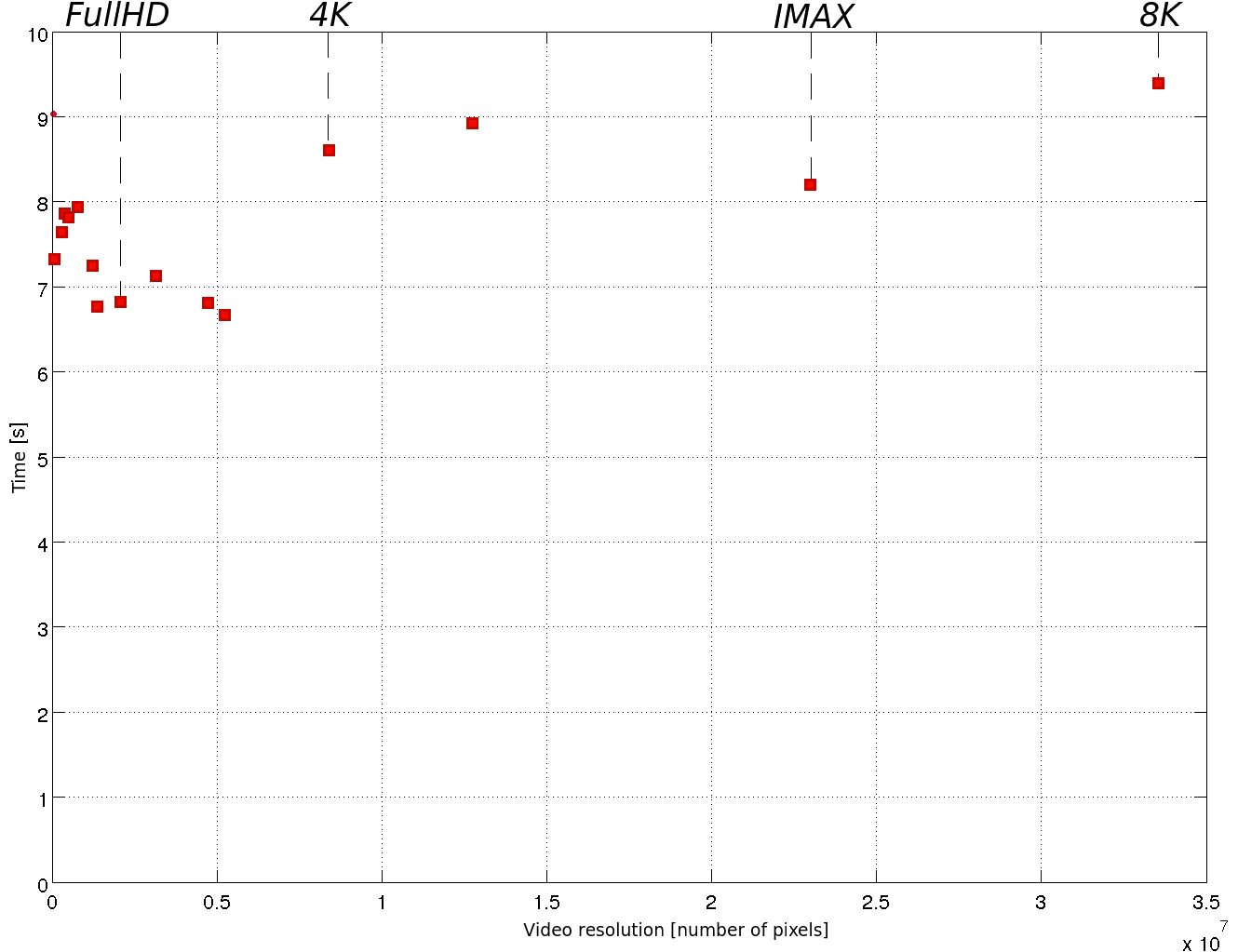}
\caption{Acceleration as a function of video resolution for \num{1000} video frames} 
\label{fig:result2}
\end{figure} 

Several experiments were conducted to determine the performance of the 
module. Fig.~\ref{fig:result1} presents the performance of both hardware and 
software implementation of the video quality assessment module for variety of 
resolutions. Starting from very low resolutions QVGA (\num{320 x 240}) and VGA 
(\num{640 x 480}), through fullHD (\num{1920 x 1080}) to UHD resolutions 4K 
(\num{4096 x 2160}) and 8K (\num{7680 x 4320}). Due to variety of display 
aspect ratios, for 4K and 8K we chose power of two values to determine the aspect 
ratio. Resulting image was around \SI{1}{\percent} larger than popular $16:9$.

The green line indicates a real-time processing performance, assuming that the 
video is streamed at \num{30} frames-per-second rate. Hardware version of the video 
quality assessment module is capable of processing 8K in real-time.

Fig.~\ref{fig:result2} presents acceleration results as a function of video 
resolution. It is worth noting that the resolution has a direct impact on the 
size of a single chunk of data sent over InputStream, which in turn affects 
transfer rate and the overall processing time.

The acceleration (Fig.~\ref{fig:result2}) is a speed-up achieved by a hardware 
solution compared to the software one. Tab.~\ref{tab:ImplementationResults} 
presents hardware resources consumption of registers (\#reg) and lookup 
tables (\#lut) in Pico platform as a function of a 
number of vqFPGA modules implemented, as well as corresponding 
throughput achieved.

\begin{table}
\caption{Hardware resources consumption (Xilinx Virtex-6 LX240T FPGA)\label{tab:ImplementationResults}}
\centering
\begin{tabular}{cccc} \toprule
\# vqFPGA modules & \#reg & \#lut & throughput [\si[per-mode=symbol]{\giga\byte\per\second}]  \\ \midrule
\num{1} & \num{23982} (\SI{7}{\percent})  & \num{11836} (\SI{6}{\percent}) & \num{0.66} \\
\num{2} & \num{29179} (\SI{10}{\percent}) & \num{15915} (\SI{7}{\percent}) & \num{1.27} \\
\num{3} & \num{34182} (\SI{12}{\percent}) & \num{19330} (\SI{9}{\percent}) & \num{1.6} \\
\num{4} & \num{39379} (\SI{15}{\percent}) & \num{23214} (\SI{11}{\percent}) & \num{1.96} \\
\num{5} & \num{44576} (\SI{18}{\percent}) & \num{27398} (\SI{13}{\percent}) & \num{2.11} \\
\num{6} & \num{49773} (\SI{20}{\percent}) & \num{31320} (\SI{14}{\percent}) & \num{2.19} \\ \bottomrule
\end{tabular}
\end{table}

\section{Summary}
Presented solution is capable of calculating simultaneously four distinct video 
quality assessment metrics on a single video stream. Used hardware platform 
allowed for real time processing of 8K resolution. Solution based on CPU only 
did not meet real time requirements for resolutions higher than fullHD. Whole 
project was implemented using Impulse~C, a high level language that 
significantly reduced design time, facilitated the system integration process 
and enabled architectural optimization which boosted the overall performance of 
the solution. Some improvements can still be done, more metrics can be added, 
also due to low resource utilization more parallel modules can be implemented 
inside the FPGA, what could further speed up calculations. If a theoretical 
highest throughput for Pico M503 platform (\SI[per-mode=symbol]{3}{\giga\byte\per\second}) will be reached, it would 
allow to process 16K resolution with \num{24} fps, which is the minimum that can be 
considered as a real time. However, because Impulse~C language allows for 
seamless moving of the design between different platforms, with Platform Support 
Package provided, presented solution can utilize more efficient hardware to 
achieve even better results.

\bibliographystyle{elsarticle-num}
\bibliography{mybibfile}

\end{document}